\documentclass{ws-ijmpa}

\begin{document}

\markboth{C.~Wilkin} {Meson Production at COSY-TOF and COSY-ANKE}

%
\catchline{}{}{}{}{}
%

\title{MESON PRODUCTION AT COSY-TOF AND COSY-ANKE}

\author{Colin Wilkin}

\address{Physics and Astronomy Department, UCL, Gower St., London WC1E 6BT, U.K.\\
cw@hep.ucl.ac.uk}
 \maketitle

\begin{history}
\received{Day Month Year}
\end{history}

\begin{abstract}
The roles of the COSY-TOF and COSY-ANKE spectrometers in the
measurement of strange meson production are briefly reviewed, mainly
in connection with new results on the $pp\to K^+p\Lambda$, $pp\to
K^+p\Sigma^0$ and $pp\to K^+n\Sigma^+$ reactions.

\keywords{Strange mesons; hyperons; cusps.}
\end{abstract}

\ccode{PACS numbers: 14.40.Aq, 14.20.Jn, 13.75.-n}

\section{Introduction}

The COSY-TOF and COSY-ANKE facilities are different in almost every
respect. The Time-of-Flight spectrometer\cite{TOFNOW} sits on an
external beam line and its barrel has a tremendously large
acceptance. It relies for its success on measuring the velocities of
many particles, with the possibility of afterwards doing kinematic
fits. It is especially useful for neutral strange particle production
because the time-delayed vertex, from say $\Lambda\to p\pi^-$, can be
shown to come from a position downstream of the target.

ANKE\cite{BAR2001}, on the other hand, is a magnetic spectrometer
situated at an internal target station of the circulating COSY beam.
It can measure the momenta of a variety of positive and negative
ejectiles, in particular $K^+$ and ``spectator'' protons. However,
the overall acceptance is generally very small and often only tiny
bits of phase space are sampled. Unlike TOF, it can be used to
perform inclusive or semi-inclusive experiments.

Despite their very different characteristics, both ANKE and TOF are
well suited for strange particle studies and we shall here
concentrate on these. In their seminal 1955 book on particle
physics\cite{BET1955}, Bethe and de Hoffmann called the $\Lambda$ and
$\Sigma$ baryons, as well as the $\theta$ and $\tau$ mesons,
``curious'' particles. It is unfortunate that this nomenclature did
not stick or we could have been saying that the $K^+$ had curiosity
value $+1$ and talk about ``associated curiosity production''!

%
%
\section{The $\boldsymbol{pp\to K^+p\Lambda}$ reaction}

The most detailed measurements of differential observables in the
exclusive $pp\to K^+p\Lambda$ reaction were carried out at the
COSY-TOF facility\cite{ABD2010,ABD2010a}. Evidence for both the
importance of the $\Lambda p$ final state interaction (FSI) and the
$N^*(1650)$ isobar is clear from the Dalitz plots obtained at 2.95,
3.20, and 3.30~GeV/$c$. Angular distributions in the overall c.m.\
frame show significant anisotropy for the $\Lambda$ and proton,
whereas the $K^+$ is fairly flat. The combined data set will provide
guidance in the construction of theoretical models but let me
concentrate here on just one aspect, viz.\ the behavior of the $K^+$
missing-mass spectra in the vicinity of the $\Sigma N$ thresholds.

\begin{figure}[hbt]
\begin{center}
\psfig{file=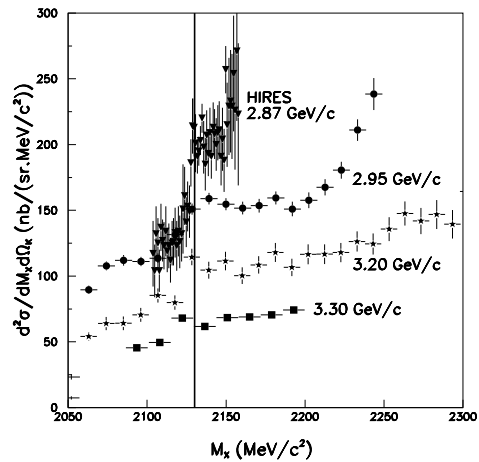,width=7cm}
\end{center}
\caption{Forward inclusive cross section for $pp\to K^+X$ as a
function of the missing-mass $M_X$. The COSY-TOF data at 2.95, 3.20,
and 3.30~GeV/$c$ were derived from \emph{exclusive} $pp\to
K^+p\Lambda$ measurements\protect\cite{ABD2010} by assuming that the
distributions in the $K^+$ c.m.\ angle are isotropic. For clarity of
presentation, the 3.30~GeV/$c$ data have been scaled by a factor of
0.6. The HIRES data at a beam momentum of 2.87~GeV/$c$ (inverted
triangles)\protect\cite{BUD2010} were directly measured in an
inclusive experiment. The vertical line indicates the position of the
average $\Sigma^+n/\Sigma^0p$ threshold.\label{money-plot}}
\end{figure}

Using the isotropy in the $K^+$ production angle\cite{ABD2010a}, we
show in Fig.~\ref{money-plot} the forward $K^+$ differential cross
section deduced from the COSY-TOF data\cite{ABD2010}. For all beam
momenta there is an enhancement at the $\Sigma N$ threshold that must
be due to the $\Sigma N \to \Lambda p$ channel coupling. This cusp
effect becomes more pronounced as the beam momentum is lowered and
this might be due to the overlap of the $N^*(1650)$ with the cusp
region then representing a larger fraction of the available phase
space.\vspace{-5mm}

Also shown in Fig.~\ref{money-plot} are the high resolution forward
inclusive $K^+$ production data from the HIRES
collaboration\cite{BUD2010}. These show a jump from the left to the
right of the $\Sigma N$ threshold which has been interpreted as being
primarily due to $\Sigma$ production\cite{BUD2010}. However, since
the exclusive COSY-TOF data show exactly the same kind of jump,
though somewhat smaller in magnitude and smeared by resolution and
binning, it is clear that a large fraction of what was assumed to be
$\Sigma$ production corresponds in reality to the $pp\to K^+p\Lambda$
reaction\cite{VAL2010a}.

%
%
\section{The $\boldsymbol{pp\to K^+n\Sigma^+}$ reaction}

Although there exist high energy bubble chamber data, the first
modern $pp\to K^+n\Sigma^+$ experiment was carried out at COSY-11 by
detecting the neutron in coincidence with the presumed
$K^+$\cite{ROZ2006}. Remarkably large cross sections were found at
excess energies of $Q=13$ and 60~MeV; these were about two orders of
magnitude higher than those for $\Sigma^0$ production. The initial
measurement at ANKE at 126~MeV gave a much lower result\cite{VAL2007}
and it seems implausible that there could be an anomalous FSI to make
these two data sets consistent. The $\Sigma^0p$ and $\Sigma^+n$ final
states are both mixtures of isospin $I=1/2$ and 3/2, and there is no
unusual $\Sigma^0p$ behavior. In order to investigate this, further
measurements were undertaken at ANKE at $Q=13$, 47, 60 and
82~MeV\cite{VAL2010} and cross sections evaluated using three
different techniques: (1) Study of inclusive $K^+$ production, (2)
Study of $K^+p$ coincidences, and (3) Study of $K^+\pi^+$
coincidences. All three methods give consistent answers and show that
the $\Sigma^+$ production cross section is just a little less than
that for $\Sigma^0$.

The most convincing of the inclusive measurements arises from looking
at the ratio of $pp\to K^+X$ data taken 13~MeV above the $\Sigma^+ n$
threshold to those 5~MeV below, since this does not depend upon the
efficiency for $K^+$ detection. The ANKE acceptance so close to
threshold is quite high and the ratio shows that, at the 95\%
confidence level, the ratio of total production cross sections
$\sigma(\Sigma^+)/\sigma(\Sigma^0) \lesssim 2$.

\begin{figure}[hbt]
\begin{center}
\psfig{file=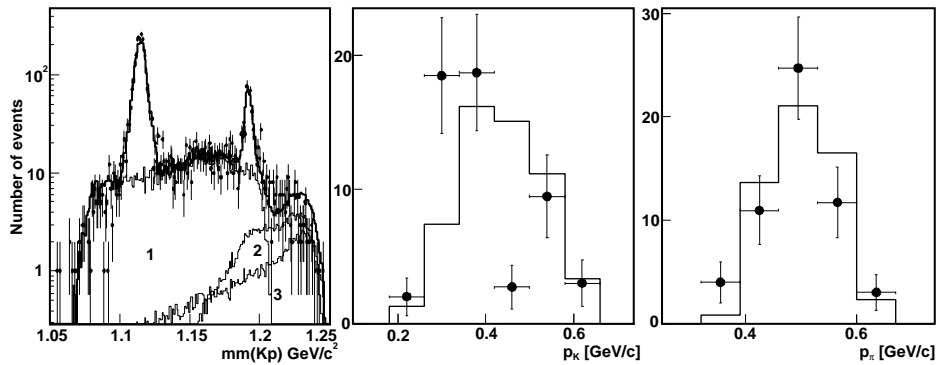,angle=-90,width=0.98\textwidth}
\end{center}
\vspace*{8pt} \caption{Left: The $K^+p$ missing-mass spectrum at
1.958~GeV\protect\cite{VAL2010}. Apart from peaks corresponding to
direct $\Lambda$ and $\Sigma^0$ production, there are also
contributions where the proton has come from the decay of a $\Lambda$
(1), a $\Sigma^0$ (2), and a $\Sigma^+$ (3). Center and Right: $K^+$
and $\pi^+$ momentum spectra from $K^+\pi^+$ coincidences at
1.958~GeV compared to simulations\protect\cite{VAL2010}.
\label{coincidences} }
\end{figure}

The results of the $K^+p$ coincidence measurements at 1.958~GeV
presented in Fig.~\ref{coincidences} show peaks corresponding to
$\Lambda$ and $\Sigma^0$ production, where the proton detected is the
``direct'' one. In addition there are indirect contributions where
the proton comes from the decay of $\Lambda,\,\Sigma^0,\,\Sigma^+$.
The simulations shown prove that, in the neighborhood of maximum
missing masses, there can be no contamination from $\Lambda$
production and decay. If we then make the drastic assumption that the
$\Sigma^0$ contributes nothing at all in this region, this gives
$\sigma(\Sigma^+)/\sigma(\Sigma^0) < 1-2$, depending upon energy.

Although only upper limits have so far been quoted, cross sections
with error bars could be derived from the inclusive and $K^+p$ data.
However, at low energies any $K^+\pi^+$ coincidences detected in $pp$
collisions must come from $\Sigma^+$ production. The $K^+$ and
$\pi^+$ momentum spectra, obtained from these gold-plated events at
1.958~GeV and shown in Fig.~\ref{coincidences}, are reasonably well
described by the simulations. There are a few random coincidences
that can be estimated from the below-threshold data. These data lead
to cross section values that are completely consistent with the other
two methods and show that $\sigma(\Sigma^+)/\sigma(\Sigma^0) \approx
0.7\pm0.1$, which is a long way below the HIRES
assumption\cite{BUD2010} and two orders below that of
COSY-11\cite{ROZ2006}.

%
%
\section{The $\boldsymbol{\Sigma N}$ final state interaction}

\begin{figure}[hbt]
\begin{center}
\psfig{file=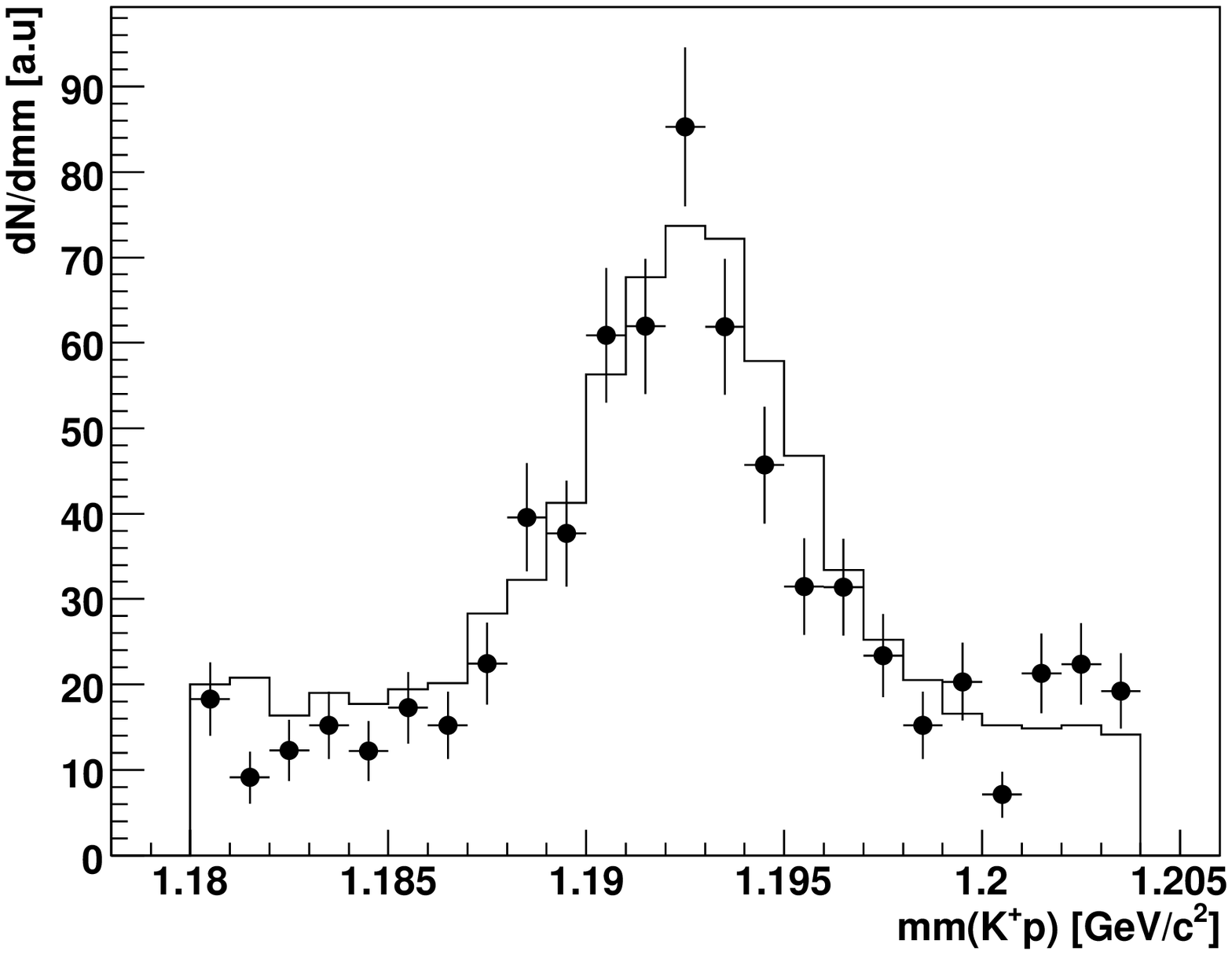,width=6.1cm} \psfig{file=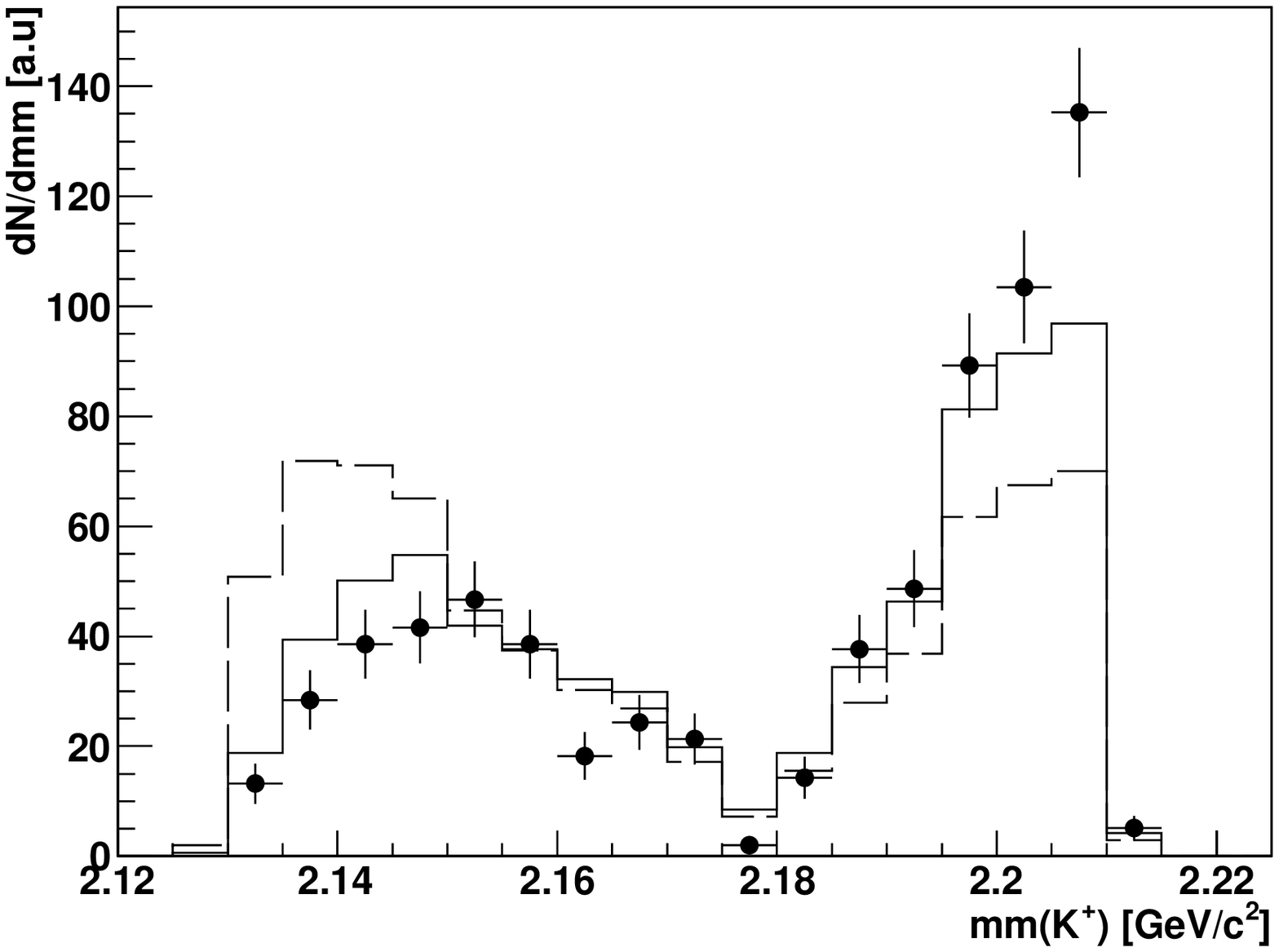,width=6.4cm}
\end{center}
\vspace*{8pt} \caption{Left: $K^+p$ missing mass from the $pp\to
K^+pX$ reaction at 2.02~GeV\protect\cite{VAL2010} showing a prominent
$\Sigma^0$ peak sitting on a physical background that is reproduced
by simulation. Right: $K^+$ missing mass, i.e.\ $\Sigma^0p$ invariant
mass, for the events in the $\Sigma^0$ peak, compared to phase-space
simulations (solid line) and these modified with a strongly
attractive $\Sigma^0p$ FSI (dashed line).\label{FSI} }
\end{figure}

Although the energy dependence of the $pp\to K^+p\Sigma^0$ total
cross section is inconsistent with a strong $\Sigma^0p$ FSI, it would
be good to have independent confirmation. The potential for this is
provided by the ANKE $pp\to K^+pX$ data\cite{VAL2010}, an example of
which is shown in Fig.~\ref{FSI}. For the events in the $\Sigma^0$
peak in the left panel, the corresponding $\Sigma^0p$ invariant mass
distribution in the right panel seems to be closer to phase space
rather than one distorted by a FSI of $\Lambda p$ strength.

Earlier experiments at COSY-TOF on $pp\to K^0p\Sigma^+$ also showed
no sign of an $I=3/2$ $\Sigma^+p$ FSI at 2.16 and 2.40~GeV, though
the FSI region was comparatively small at these higher
energies\cite{KAR2005}.
%
%
\section{Hyperon production in proton-neutron collisions}

All studies on the neutron have so far been undertaken in quasi-free
kinematics using a deuterium target. The $pd \to p_{\rm
sp}K^0p\Lambda$ reaction has been investigated at
COSY-TOF\cite{KRA2009} by detecting the combination of \emph{two}
delayed decays, viz.\ $K^0 \to \pi^+\pi^-$ and $\Lambda\to p\pi^-$,
for which the apparatus is ideal. The momentum of the spectator
proton, $p_{\rm sp}$, was then deduced through a kinematic fit. About
1000 fully reconstructed events are expected from the pilot run,
though the c.m.\ energy has to be evaluated event-by-event and so the
data are spread over a wide range.

A complementary approach has been proposed at ANKE\cite{DZY2010},
where the spectator from $pd \to p_{\rm sp}K^+n\Lambda$ will be
detected in telescopes placed in the target chamber. The spectator
determines the c.m.\ energy and, below the $\Sigma$ thresholds, the
$K^+$ signals the production of a $\Lambda$.

It should be noted that the total cross sections for $pp\to
K^+p\Lambda$, $pn\to K^0p\Lambda$, and $pn\to K^+n\Lambda$ are linked
by isospin, though there can be interferences between isospin-0 and
-1 in differential observables.

Data with direct spectator detection have already been taken at ANKE
on $pd \to p_{\rm sp}K^+pX^–$\cite{SHI2009}. The $p_{\rm sp}K^+p$
triple coincidence measurements resulted in about 500 $\Sigma^-$
events at both 2.12 and 2.22~GeV.
%
%
\section{Comparison of $\boldsymbol{pp \to pK^+\Sigma^0\pi^0}$
and $\boldsymbol{pp \to pK^+pK^-}$ reactions}

There have been measurements of both the $pp \to
pK^+\Sigma^0\pi^0$\cite{ZYC2008} and $pp \to pK^+pK^-$\cite{MAE2008}
reactions at 2.83~GeV at ANKE. Could these two associated production
reactions be connected in some way? This might happen if the
$\Sigma^0\pi^0$  and $pK^-$ were both produced through the decay of
the $\Lambda(1405)$ resonance. What we would then be seeing is
different manifestations of the reaction $pp \to pK^+\Lambda(1405)$.

\begin{figure}[hbt]
\begin{center}
\psfig{file=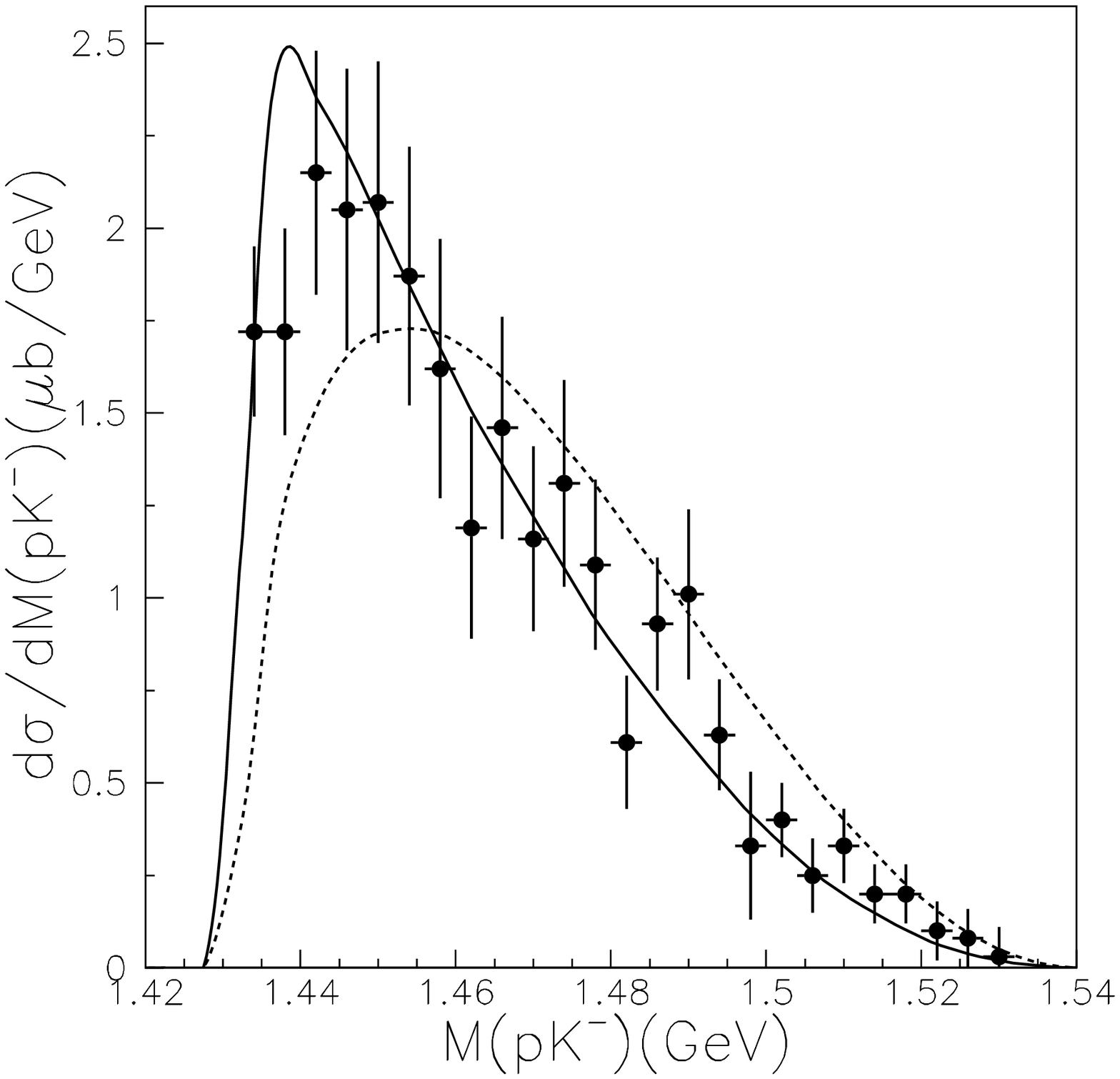,width=6.2cm} \psfig{file=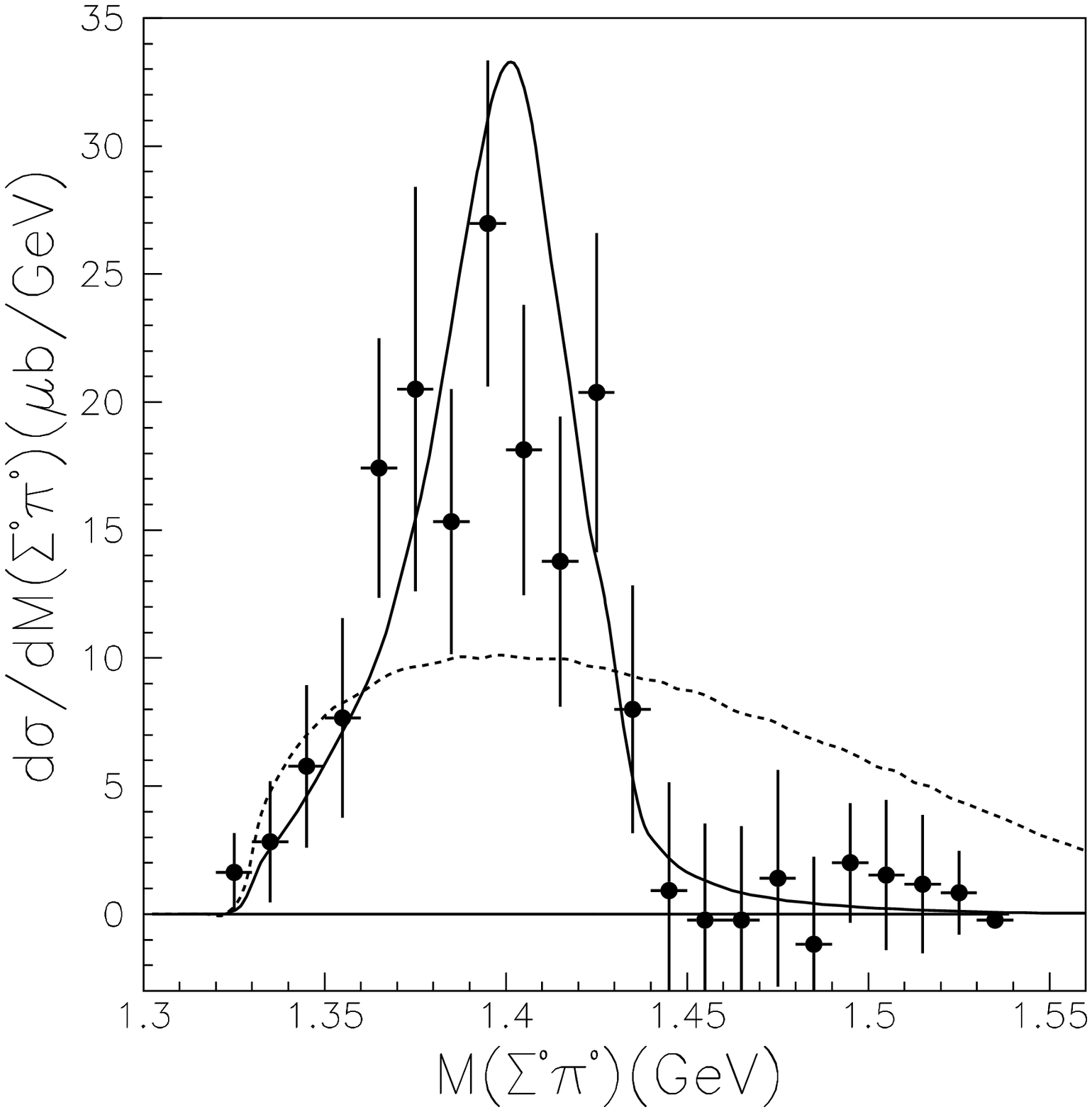,width=6.2cm}
\end{center}
\vspace*{8pt} \caption{Differential cross section for (Left) $pp \to
pK^+ K^-p$ at $Q= 108$~MeV\protect\cite{MAE2008} and (Right) $pp \to
pK^+ \pi^0 \Sigma^0$ at $Q = 212$~MeV\protect\cite{ZYC2008}. The
predictions of the $N^*(1535)$ model\protect\cite{XIE2010} are shown
by the solid lines which, in the $\Sigma^0\pi^0$ case, has been
scaled down by a factor of 2.6. The dashed lines represent normalized
four-body phase-space distributions. \label{JuJun}}
\end{figure}

In a new phenomenological meson-exchange model\cite{XIE2010}, the
reaction is assumed to proceed via the production and decay of the
$N^*(1535) \to K^+\Lambda(1405)$, since it is believed that the
$N^*(1535)$ isobar contains a lot of hidden strangeness. This
assumption determines the strengths of the cross sections but not the
shapes of the spectra shown in Fig.~\ref{JuJun}. These and the
relative strengths are fixed mainly by low energy
$\bar{K}p/\pi\Sigma$ data, which have been parametrized in terms of a
separable potential\cite{SCH2007}. The $K^-p$ predictions come out
too high by about a factor of 2.6, whereas the $\Sigma^0\pi^0$ are
about right.

This factor of 2.6 between the quality of the two predictions depends
very little at all on the $N^*(1535)$ assumptions; the effects of the
low-energy coupled-channel physics dominate. Given the theoretical
and experimental uncertainties, the plausible agreement here suggests
that $K^+K^-$ production in $pp$ collisions near threshold is also
driven mainly by intermediate hyperon states and not, for example, by
the production of scalar mesons.

In the time available I have chosen to highlight the COSY-TOF and
COSY-ANKE contributions in ``curiosity'' production, they have also
undertaken lots of non-strangeness studies. This talk must be for
another day!

\section*{Acknowledgments}
Useful information on the COSY-TOF and COSY-ANKE experiments was
supplied by W.~Eyrich and Yu.~Valdau. respectively. Support from the
conference organizers is gratefully acknowledged.

%
%

\end{document}